# Auction-based approximate algorithm for Grid system scheduling under resource provider strategies

Arash Raeisi Gahrouei, Mehdi Ghatee
Department of Computer Science, Amirkabir University of Technology, Tehran, Iran

**Abstract**
*In this paper a new mathematical model is proposed for task scheduling and resource allocation in Grid systems. In this novel model, load balancing, starvation prevention and failing strategies are stated as the constraints and the solution is restricted with a predefined quality of service for users with different priorities. These strategies are defined by resource providers based on the amount of submitted jobs to Grid. To solve the proposed model, a modern approximate Auction-based algorithm is developed and it is implemented as a prototype of Grid simulator namely Multi-S-Grid. The results are illustrated on 18 different large-scale Grid systems with different random capabilities and different users. The outcomes reveal the reasonable performance of the proposed Auction-based algorithm to solve Grid system optimization models.*

## 1. Introduction

Grid system is a federation of computer resources from multiple administrative domains which are shared among users by establishing a global resource management architecture [1]. The management of a Grid system requires both allocation and coordination resources to execute both user's jobs assignment to resources and data management through the network. The jobs request Grid resources dependently or independently. The resources allocation system assigns requirements to resources based on Grid scheduler strategies. The scheduler usually consists of two hierarchical problems in two different levels; global-scheduling and local-scheduling. The global-scheduler receives jobs and selects the resources for a job. In the local-scheduling level, the assigned jobs are scheduled in each resource [2]. A global-scheduler can be developed in centralized or distributed versions. In the first version, based on colleting all users' requests, recourse utilization information and overall knowledge of the available recourses, a central global-scheduler is used to assign jobs to resources [3,4,5]. In the distributed version, a number of distributed global-schedulers keep track of the available resources and schedule the jobs. In [2] the centralized and decentralized designs were compared. Also in [6] a novel distributed design was given. Besides in [7], the centralized, the distributed, and the hierarchical job-scheduling processes have been discussed.

For job scheduling in Grid systems, several algorithms have been proposed including FPLTF (Fastest Processor to Largest Job First) [8,9], WQR (Work Queue with Replication) [9], Min-min and Max-min[10,11]. On the other hand, global-heuristics algorithms are implemented for scheduling jobs. As some instances, one can note to [12,13] including genetic algorithm, [12] consisting of simulated annealing, [14,15] considering ant colony optimization, and [16] applying multi-objective evolutionary algorithm.

Because of the importance of Grid scheduling, some researchers have focused on Grid systems models and simulation. For example, Buyya and Murshed [17] developed a Java-based discrete-event Grid simulation toolkit called GridSim and evaluated the performance of deadline and budget constraints. Then the necessary components for GridSim were developed, see e.g., [18].

However this framework cannot be simply generalized using state-of-the-art algorithms. Instead of such generalization, we implement a new simulator namely multiple strategies Grid or shortly Multi-S-Grid to schedule the jobs in a balanced plan. Note that, resource balancing has been followed by Jain and Singh [14]. Similarly, we try to check different strategies on Multi-S-Grid. Some of the dynamic strategies have been discussed in [19, 20].

In Multi-S-Grid, a new algorithm is proposed to find the optimal or near-optimal solutions for Grid scheduling problem. The proposed algorithm is based on Auction algorithm which is a primal-dual algorithm for solving assignment problems [21]. The extensions of Auction algorithm have been given in [22]. Freling et al. [23] have also used Auction algorithm for single depot vehicle scheduling problem. Attanasio et al. [24] used Auction algorithms for decentralized parallel machine scheduling systems. Das and Grosu [25] also introduced the approximation Auction algorithm for resource management in Grids and they studied three types of auction allocation protocols including first-price auction, Vickrey auction and double auction for resource management in [26], which are different from traditional Auction algorithm. Also note that, the auction option in GridSim, including Dutch-auction and English-auction, is not adopted from famous Auction algorithm in [21] and it is a greedy algorithm. However in Multi-S-Grid a hybrid version of the approximate algorithm of [21] is completely developed. In other words, Auction algorithm in Multi-S-Grid can be simplified to the auction algorithm in GridSim when only a single job exists. In other cases, the contribution of this paper improves the scheduling results.

In the remainder of this paper, a new mathematical model for the Grid scheduling problem is given in Section 2. Section 3 provides Auction based approximate algorithms for solving this problem. Multi-S-Grid which is a new Grid simulator is described in Section 4 and the results of simulation are illustrated. Section 5 ends this paper with a brief conclusion and future remarks.

## 2. **Mathematical model of Grid resource management**

A Grid system consists of four main components: users, resources, resource manager and Grid Information System (GIS). In what follows, we use S, MI and KB instead of "Second", "Million Instructions" and "Kilobyte".

A user has several jobs. Job $j$ has several characteristics: job identifier denoted with ID, job owner, arrival time denoted with $at_j(S)$, priority denoted with $p_j$, length denoted with $l_j(MI)$ and volume denoted with $v_j(KB)$. Also user $i$ has several characteristics including: user identifier denoted with ID, requested quality of service denoted with $qos_i$, user network bandwidth denoted with $bw_i(\frac{KB}{S})$ and network fail rate denoted with $bf_i(S)$.

A resource $r$ consists of $M_r$ machines. Machine $m$ includes machine identifier denoted with ID, similar processing quality denoted with $pq_m(\frac{MI}{S})$ and $c_m$ similar processors. Each machine may be failed with a fail rate $f_m(S)$. A resource also consists of: network bandwidth denoted with $bw_r(\frac{KB}{S})$, the network fail rate denoted with $bf_r(S)$ and processing quality denoted with $pq_r(\frac{MI}{S})$. $pq_r$ is the average of processing qualities of free available processors in each machine of resource $r$. Note that if a machine of a resource is failed, only this machine will be unreachable, but if network of resources is failed, all of its machines will be unreachable.

Resource manager consists of a global scheduler and several local schedulers. When a job is submitted to Grid system by owner in time $t$, the global scheduler puts it in a waiting list (global list) with respect to jobs' characteristics. According to different goals, the jobs of global list are assigned to resources and they are eliminated from global list. For a resource, local scheduler submits the assigned jobs to its free processors for executing. When a job executes, the result is forwarded to owner. Assume that the jobs should be performed in a preemptive manner, thus once a job executes on a processor, it cannot be terminated up to completion.

In Figure 1, the dependency between the entities in the described Grid is presented. Such conceptual model is implemented in this paper to analyze the different strategies. In what follows, the mathematical models of GIS, global scheduler and local schedulers are stated.

/**** Location of Figure 1 ****/

### 2.1. GIS modeling

GIS saves the information about all resources and the assigned jobs and updates the characteristics of the system components before each of the schedulers' executions. Let $R(t), J(t), U(t)$ and $M(t)$ be the sets of resources, jobs, users and machines up to time $t$. To model GIS, the available variables can be stated as follows:

$$S_i(t) = \begin{cases} 1 & \text{if } i \text{ exists in the system in time } t, \\ 0 & \text{otherwise,} \end{cases}$$

where $i \in R(t) \cup J(t) \cup U(t) \cup M(t)$.

We have the following rules in Grid system:

If $S_i(t) = 0, for\ i \in U(t)$, then i$^{th}$ user's jobs are removed from system.
If $S_i(t) = 0, \ for\ i \in J(t)$, then job i is removed from the system.
If $S_i(t) = 0, \ for\ i \in R(t)\ and\ i \in M(t)$, then all of the jobs serviced by resource $i$ (machine i), are removed from system.

Now, define the following assignment variables:

$$x_{j_i r}(t) = \begin{cases} 1 & \text{if job } j \text{ of user } i \text{ is assigned to resource } r \text{ at time } t, \\ 0 & \text{otherwise,} \end{cases}$$

$$x_{j_i m}(t) = \begin{cases} 1 & \text{if job } j \text{ of user } i \text{ is assigned to machine } m \text{ at time } t, \\ 0 & \text{otherwise.} \end{cases}$$

$Global\_Jobs(t)$ is the set of unassigned jobs at time t and can be stated as follows:

(1)
$$\begin{cases} Global\_Jobs(0) = \emptyset \\ Global\_Jobs(t) = \{j_i \in Global\_Jobs(t-1)\ |S_i(t) = 1, S_j(t) = 1, x_{j_i r}(t) = 0\} \\ \qquad\qquad \cup \{j_i \in J | at_{j_i} = t\} \end{cases}$$

where $j_i$ is a job of user $i$.

The number of free processors on resource r at time t is given by the following equation:

(2) $$ACR_r(t) = \left( \sum_{m \in M_r}(c_m . S_m(t)) - \sum_{l=0}^{t} \sum_{j_i \in J} \left( x_{j_i r}(t) . f(et_{j_i} - t) \right) \right) . S_r(t)$$
$$\forall r \in R(t)$$

where $c_m$ is the number of processors of machine m, $et_{j_i}$ is the termination time of job $j_i$ and $f$ is a Heaviside step function which can be stated as follows:

$$f(x) = \begin{cases} 1 & x < 0 \\ 0 & x \geq 0 \end{cases}$$

Note that the set of available resources at time t which is denoted with $Ready\ Resources(t)$, can be defined as follows:

(3) $\quad Ready\_Resources(t) = \{r \in R(t)\ |\ ACR_r(t) > 0\}$

The number of free processors on machine m at time t is given by the following equation:

(4) $$ACM_m(t) = \left( c_m - \sum_{l=0}^{t} \sum_{j_i \in J} \left( x_{j_i m}(t) . f(et_{j_i} - t) \right) \right) . S_m(t) \qquad \forall m \in M(t)$$

Note that the set of available machines on resource r at time t which is denoted with $Ready\_Machines_r(t)$, can be given as bellow:

(5) $\quad Ready\_Machines_r(t) = \{m \in M_r \mid ACM_m(t) > 0\}$

$Local\_Jobs_r(t)$ is the set of jobs which are assigned to the resource r at time t and it is defined as the following:

(6) $\quad Local\_Jobs_r(t) = \{J_i \in Global\_Jobs \mid x_{j_ir}(t) = 1\}$

In what follows, based on the proposed entities, the mathematical models of Grid schedulers are stated.

## 2.2. Schedulers modeling

There are different models for Grid scheduling problem, see e.g., Fibich et al. [0]. In these models different distributed structures, various jobs, and several optimality criteria are considered. In this paper, the following model is defined for global scheduler at time t:

(7) $\quad Min \sum_{r \in Ready\_Resources(t)} \sum_{j_i \in Global\_Jobs(t)} c_{j_ir}(t) \cdot x_{j_ir}(t) \quad \forall t$

s.t.

(8) $\quad \sum_{j_i \in Global\_Jobs(t)} x_{j_ir}(t) \leq ACR_r \quad \forall r \in Ready\_Resources(t), \forall t$

(9) $\quad \sum_{r \in Ready\_Resources(t)} x_{j_ir}(t) \leq 1 \quad \forall j_i \in Global\_Jobs(t), \forall t$

(10) $\quad x_{j_ir}(t) \in \{0,1\}$

In the objective function (11), $c_{j_ir}(t)$ is the sum of transfer time of job $j_i$ to resource r at time t and the processing time of job $j_i$ on resource r at time t. Thus $c_{j_ir}(t)$ can be defined as the following, see [14, 27]:

(11) $\quad c_{j_ir}(t) = \frac{v_{j_i}}{\min(bw_i, bw_r)} + \frac{l_{j_i}}{pq_r(t)}$

in which $pq_r(t)$ is processing quality of resource $r$ and can be given with:

(12) $\quad pq_r(t) = \frac{\sum_{m \in M_r}(ACM_m(t) \times pq_m)}{\sum_{m \in M_r} ACM_m(t)}$.

The first term of $c_{j_ir}(t)$ is the transfer time for job j of user i. The second part depends on processing time at time t. Thus, the objective function (7) minimizes the total transfer and processing time at time t. The first constraint (8) states that the total number of assignment is less than the number of free processors on each resource and the second constraint (9) (10) ensures that each job should be assigned at most once.

When the global scheduler assigns jobs to resources, each resource uses a local scheduler to assign jobs to machines. The model of local scheduler is stated as follows:

(13) $\quad Min \sum_{m \in Ready\_Machines_r(t)} \sum_{j_i \in Local\_Jobs_r(t)} c_{j_im} \cdot x_{j_im}(t) \quad \forall t$

s.t.

(14) $\quad \sum_{j_i \in Local\_Jobs_r(t)} x_{j_im}(t) \leq ACM_m(t)$
$\quad\quad\quad \forall m \in Ready\_Machines_r(t), \forall r \in R(t), \forall t$

(15) $\quad \sum_{m \in Ready\_Machines_r(t)} x_{j_im}(t) = 1 \quad \forall j_i \in Local\_Jobs_r(t), \forall t$

(16) $\quad x_{j_im}(t) \in \{0,1\}$

In this model $c_{j_im}$ is the processing time of job $j_i$ on the machine $m$ of resource $r$ and is defined as $c_{j_im} = \frac{l_{j_i}}{pq_m}$ in which $l_{j_i}$ and $pq_m$ are the length of job $j$ and the processing quality of machine m, respectively. The first constraint implies that the total number of assignment is less than the number of free processors on each machine of resources and the second constraint states that each job should be assigned to a unique processor.

## 2.3. Grid management strategies

The different strategies in Grid management systems has been presented in [20] as an instance. In pursuing this work, the following strategies are considered in the proposed Grid systems:

- Starvation prevention: Starvation of a job is defined as too long waiting time for a job in global list because sometimes the time of transfer and processing of a job is very long or the free processor is not available for a long time.
- Prevention of job failing: The processing of a job may be failed because of a failure in a machine after job accepting. Also the network link between owner of a job and its resource can be failed. In these cases, the fail rate should be minimized.
- User's quality of service observance: Each user requests a level of quality of service. Higher quality of service decreases the processing time of jobs, so the total number of user's job will be increased. Resource provider can be supposed as a user with a high quality of service such that if resource provider submits a management job to Grid system, the job will be processed as soon as possible.

In this paper, three parameters ($fp, qp$ and $sp$) are defined by resource provider to set requirements. These parameters are used for fail prevention, quality of service and starvation prevention, respectively.

Load balancing observance of resources and machines: The Grid scheduler is designed to balance the load on the global level and on the local level if it is required. In the global level, the load balancing is performed for resources while in the local level the load for different machines can be balanced. The load balancing constraint in the global level is stated as follows:

$$(17) \quad \frac{ACR_r(t)}{\sum_{m \in M_r}(c_m \cdot S_m(t))} \cong \frac{ACR_{r'}(t)}{\sum_{m \in M_{r'}}(c_m \cdot S_m(t))} \quad \forall r, r' \in Ready\_Resources(t), \forall t$$

In the next section, we discuss on the application of this linear system to balance the loading of resources in Auction-based approximate algorithm for Grid global scheduling.

Also the load balancing constraint in the local scheduler is considered on the loads of machines. This constraint is stated as follows:

$$(18) \quad \frac{ACM_m(t)}{(c_m \cdot S_m(t))} \cong \frac{ACM_{m'}(t)}{(c_{m'} \cdot S_{m'}(t))}$$

$$\forall m, m' \in Ready\_Machines_r(t), \forall r \in Ready\_Resources(t), \forall t$$

This linear system will be used in Shortest Job First (SJF) for Grid local scheduling in the next section.

## 3. Auction based approximate algorithms for Grid scheduling

In this section the Auction algorithm for assignment problem is reviewed. Then, a new heuristic algorithm based on the Auction algorithm is proposed for resource management. The Auction algorithm is one of the most powerful and efficient algorithm for different kinds of assignment problems [21].

### 3.1. The Auction algorithm for the assignment problem

The simplest assignment problem is to match $n$ persons to $m$ objects on a one-to-one relation. The benefit of matching a person $i$ to an object $j$ denotes with $a_{ij}$. The objective is to assign persons to objects in a way to maximize the total benefit. The set of objects which can be assigned to person $i$ is $A(i)$. An assignment $S$ is a set of person-object pairs $(i, j)$ that $j \in A(i)$. An assignment is said to be feasible if it contains $n$ pairs, so that every person and every object is assigned; otherwise the assignment is called partial, see [21].

The Auction algorithm proceeds iteratively and terminates when a feasible assignment is obtained. At the start of the algorithm we have a partial assignment $S$ and a price vector $p$ satisfying $\varepsilon$-complementary slackness ($\varepsilon - CS$). The $\varepsilon - CS$ conditions for the assignment problem is stated as follows [21]:

(19) $\quad a_{ij} - p_j \geq \max_{k \in (Ai)} \{a_{ik} - p_k\} - \varepsilon \quad \forall (i,j) \in S$

As an initial choice, one can use an arbitrary set of prices together with the empty assignment. The iteration consists of two phases: the bidding phase and the assignment phase, see [21] for details and description. The worst-case running time of the Auction algorithm is $O(n|A| \log(nC))$ $O(n|A| \log(nC))$ in which $C = \max_{(i,j) \in A} |a_{ij}|$ see [21].

### 3.2. Auction based heuristic algorithm for Grid scheduling

The model of global and local scheduler can be represented as an assignment problem in which the demand of each job is one and the supply of each resource (machine) is equal to its number of free processors. If there are m jobs in Global_Jobs(t) and the number of free processors is n at time t, then in the first step the scheduler tries to equalize n and m by adding some virtual processors or some virtual jobs. So we can consider the following assignment model for the global scheduler:

(20) $\quad \min \sum_{(j,r) \in A} a_{jr} x_{jr}$

s.t

(21) $\quad \sum_{\{r|(j,r) \in A\}} x_{jr} = 1 \quad \forall j = 1 \dots m$
(22) $\quad \sum_{\{r|(j,r) \in A\}} x_{jr} = B_r \quad \forall r = 1 \dots n$
(23) $\quad \sum_{r=1}^{n} B_r = m \quad \forall r = 1 \dots n$
(24) $\quad x_{jr} = \{0,1\}$

In this model $m$ is the number of unassigned jobs and $B_r$ is the number of free processors of recourse $r$. $a_{jr}$ is the processing cost of job $j$ on recourse $r$. The $\varepsilon - CS$ conditions for the assignment model can be defined as follows [21]:

(25) $\quad p_j - p_r \leq a_{jr} + \varepsilon \quad \forall (j,r) \in A, (x_{jr} < 1)$
(26) $\quad p_j - p_r \geq a_{jr} - \varepsilon \quad \forall (j,r) \in A, (x_{jr} > 0)$

Based on the $\varepsilon - CS$ conditions, a powerful heuristic can be proposed for global scheduler, see the flowchart presented in Figure 2 whose structure is adopted from [21]. Note that as mentioned in Section 2, the cost function $a_{jr}$ has enough potential for implementing different strategies and constraints. By this parameter it is easy to define a weighted combination of different strategies requested by users and resource providers. Also by $g_r$ which can be obtained by solving a linear system of balancing constraints given in (17) or the maximum available capacity of resources given in (2), it is possible to balance the loads on different resources.

/**** Location of Figure 2 ****/

In the local level assignment, a job with the lowest processing time is assigned to a machine with the highest processing quality. Note that the complexity of this algorithm is polynomial because of similarity with that of presented in [21] for assignment problem.

## 4. Grid simulator description

In this section we describe a new Grid simulator to model online and dynamic scheduling and we investigate the effects of different strategies on Grid systems according to different values of $ff, qf$ and $sf$p and *sp* parameters.

### 4.1. The Grid simulator

To present the details of Multi-S-Grid, its components are described in Figure 3. As one can note that the users' files and jobs files are sent to jobs pool. After assigning the jobs to resources by global scheduler, the jobs pool sends the jobs to resources and in this time, the local scheduler assigns the jobs to processors. All of the information is saved in GIS. When the jobs are assigned, normally terminated or probably failed, the report system saves the jobs statues and GIS updates the resources statues.

/**** Location of Figure 3  ****/

### 4.2. Input functions

- *Grid function*

Based on the parameters of Grid system, Multi-S-Grid is developed. The parameters are as follows: the number of recourses, their minimum and maximum bandwidth level, the minimum and maximum of fail rate of network links, minimum number and maximum number of machines in each resource, the specific properties and characteristics of machines such as fail rate of machines, minimum number and maximum number of processors in each machine, minimum and maximum of processing quality.

Grid management system defines a function applying some internal functions to specify the Grid system and save it in a distinct file. The sample of this file is illustrated in Figure 4.

/**** Location of Figure 4  ****/

- *User function*

This function saves a file including user's information such as the number of users, their bandwidth, the level of fail rate and their quality of service level. The sample of this file is illustrated in Figure 5.

/**** Location of Figure 5  ****/

- *Job function*

This function saves the characteristics of jobs such as the owner of job, required processing time, volume and priority of job. The sample of this file is illustrated in Figure 6.

/**** Location of Figure 6  ****/

Multi-S-Grid receives the files of Grid system, users and jobs and organizes these files. It consists of several parts which are presented in the following.

### 4.3. Resource prepared function

This function updates the list of available resources of the Grid system to tune up the parameters of global scheduler for assigning the next jobs. This system receives the failing reports from resources and machines and sends a flag for GIS. With respect to these flags, the GIS passes the failed jobs to the report system.

### 4.4. GIS Implementation

The GIS controls the status of assigned jobs. For this aim, the GIS is implemented as an array with several records about the assignments including job ID, recourse ID, machine ID, arrival time, assignment time, transfer time and processing time. When a job is assigned or the processing time of a job is finished the GIS updates the information.

### 4.5. Report system

This system saves the information of terminated jobs with its statues such as those which are failed by user and also failing resource or machine, etc. When the processing of a job is completed or is failed, its record is deleted from the GIS and added to the report system. This system is used for analyzing the Grid system.

### 4.6. Data preparing system

This system is the most important component of the simulator. The system designs the cost of service such that the objective of Grid system with different strategies is complied. In this paper, we use the expecting time which can be modeled as the following:

(27)   (1)   $a_{j_i r} = \left(pt_{j_i}\right)$

in which $a_{j_i r}$ is expecting time of job j on resource r, $pt_{j_i r}$ and $st_{j_i r}$ are the transfer and processing time of job j on recourse r, respectively and f is defined by the following relation:

(28)   $f = \frac{(p(pt_{j_i r}) \times (p(st_{j_i r}))^{fp}}{qos_i^{qp} \times \rho_{j_i}^{sp}},$

where $p(pt_{j_i r}) = e^{-\frac{pt_{j_i r}}{af_r}}$ is the probability of no fail in processing of job, $af_r$ is the average fail rate of recourse r. $p(st_{j_i r}) = e^{-\frac{st_{j_i r}}{bf_r}}$ or $p(st_{j_i r}) = e^{-\frac{pt_{j_i r}}{bf_i}}$ is the probability of no fail in transferring of job. $bf_r$ and $bf_i$ are the fail rate for recourse and user connection, respectively. $fp, sp$ and $qp$ are the control parameters and sets by resource provider to meet own goals.

## 5. Simulation results

To simulate different strategies in Grid scheduling, in this paper a prototype of a new Grid simulator namely Multi-S-Grid is developed on MATLAB engineering software. The structure of this simulator is illustrated in Figure 3. We consider 18 Grid systems with different properties and two groups of different users. Because of paper limitation, we chose 10 Grids between these 18 Grids randomly. For these experiments, the Grid properties and the users' characteristics are presented in Table (1) and Table (2), respectively. Each user contains 30 sets of jobs with different characteristics. The Grid system is implemented with different strategies. The results of these implementations are illustrated for analyzing the effect of different parameters including starvation, fail rate, quality of service and load balancing.

/**** Location of Table 1 ****/
/**** Location of Table 2 ****/

### 5.1. Starvation parameter

Resource manager prevents from job starvation by using preferences. The parameter $sp$ is used to control the effect of starvation. The results are shown in Figure 7, Figure 8 and Figure 9. Note that the cost of a job is defined as the sum of the transfer time and the processing time. Figure 7 shows that by increasing the starvation parameter ($sf$), the number of processed job decreases. So the scheduler system assigns jobs with great cost. Figure 8 emphasizes that by increasing the parameter $sf$, the average of the costs of assigned jobs up to this step increases. Also Figure 9 shows that by increasing the parameter $sf$ the termination time of jobs increases, because the Grid system processes jobs with great costs. Also these figures show that increasing $sf$ in the interval [0, 1.5] causes to prevent from starvation, while in the interval [1.5, 2] the increasing is not beneficial.

/**** Location of Figure 7  ****/
/**** Location of Figure 8  ****/
/**** Location of Figure 9  ****/

Because starvation parameter has an important role in the proposed algorithm and the algorithm increases the priority of a job when it comes to queue, the assigned cost decreases and the probability of job assignment in next iterations increases. Thus the algorithm tries to complete the jobs with great costs.

### 5.2. Failing parameter

In this subsection we analyze the recourse fail rate, machine fail rate and the user fail rate. Assume that the failing happens in random time with Poisson distribution. Figure 10 shows the effect of failing parameter on the number of failed jobs. Figure 10 shows that by increasing failing parameter (fp), the number of failed jobs decreases, because the scheduler tries to submit jobs to resources whose probability of availability are greatest at time t.

/**** Location of Figure 10  ****/

### 5.3. Quality of service parameter

For analyzing quality of service parameter ($qp$), an experiment is implemented on 20 sets of jobs and on 18 Grid systems. The results are shown in Figure 11. This figure shows that by increasing parameter $qf$, the number of processed jobs for users with a maximum quality of service increases while the number of the processed jobs for users with minimum quality of service decreases. Also when $qf \geq 2$, the variation is constant.

/**** Location of Figure 11  ****/

### 5.4. Load balancing parameter

In this experiment, we study the effect of load balancing strategy on the resource manager. In this paper Grid is studied in two different states:
- The Grid in peak period where the number of jobs is usually greater than the number of free processors.
- The Grid under peak period where the number of jobs is usually less than the number of free processors.

Figure 12 shows the effect of load balancing parameter in a Grid with seven resources which are under peak. This figure shows that the percent of resource loading is similar for different resources when the global scheduler uses load balancing strategy.

/**** Location of Figure 12  ****/

To clarify the results, in Figure 13, we investigate the effect of load balancing parameter in two random resources of Grid which were previously presented in Figure 12. As one can note that, the loads of two resources are similar because the scheduler tries to balance loading on all of the resources.
/**** Location of Figure 13  ****/
When the load balancing strategy is not considered, the results of Figure 14 and Figure 15 are obtained. In this case the load of resource 4 is greater than that of resource 5 because the processing quality of resource 4 is greater than the processing quality of resource 5.
/**** Location of Figure 14  ****/
/**** Location of Figure 15  ****/
Figure 16 compares the variance of resource loading with and without load balancing strategy in a Grid under peak period. This figure shows that the variance of resource loading is close to zero when load balancing strategy is implied. Thus, the loading levels on the resources are balanced.
/**** Location of Figure 16  ****/
Figure 17 compares the variance of resource loading with load balancing strategy for a Grid in peak period. In this examination, after $75^{th}$ time slice, the number of submitted jobs is greater than the number of processors. This figure shows that when the Grid is under peak period, the load balancing strategy is helpful, however for Grid in peak, the effect of load balancing strategy is not noticeable, because the load of resources are full.
/**** Location of Figure 17  ****/
The analyzing of the load balancing strategy for machines provides similar results and to summarize, we ignore from presenting details.

## 6. Conclusion and Future Directions

In this paper, two new mathematical models for local and global schedulers in Grid systems are studied. We develop a structure for implementing different strategies for Grids management. The considered strategies are starvation prevention, quality of service satisfaction and resource load balancing. We implement a new Grid scheduler system namely Multi-S-Grid based on these models. Then we propose an approximate algorithm based on Auction algorithm for job assignment in Grid scheduler. The results of Multi-S-Grid are investigated on 18 Grid systems with different properties and different users. The simulation results show that the proposed model is efficient for Grid scheduling and job assignment.

Table (1) Some of the different Grids with their properties used for simulation with Multi-S-Grid (The abbreviation in Column 1 is based on the corresponding Grid properties mentioned in Figure 4)

|              | G1   | G2   | G3   | G5   | G7   | G10  | G11  | G12  | G16  | G18  |
|--------------|------|------|------|------|------|------|------|------|------|------|
| n. r.        | 3    | 5    | 7    | 5    | 3    | 5    | 7    | 3    | 5    | 7    |
| min r. b.    | 32   | 32   | 64   | 32   | 64   | 32   | 64   | 64   | 32   | 32   |
| max r. b.    | 512  | 512  | 1024 | 512  | 1024 | 512  | 1024 | 1024 | 512  | 512  |
| min r. b. f. | 30   | 15   | 30   | 30   | 15   | 15   | 15   | 15   | 15   | 30   |
| max r. b. f. | 120  | 90   | 120  | 120  | 90   | 90   | 90   | 90   | 90   | 120  |
| min n. m.    | 1    | 1    | 1    | 1    | 1    | 1    | 1    | 1    | 1    | 1    |
| max n. m.    | 4    | 8    | 4    | 8    | 8    | 4    | 8    | 4    | 8    | 8    |
| min m. f.    | 15   | 10   | 10   | 15   | 15   | 15   | 15   | 10   | 10   | 15   |
| max m. f.    | 90   | 60   | 60   | 90   | 90   | 90   | 90   | 60   | 60   | 90   |
| min p. s.    | 1200 | 2400 | 1200 | 1200 | 2400 | 1200 | 1200 | 1200 | 1200 | 1200 |
| max p. s.    | 3600 | 3600 | 3600 | 3600 | 3600 | 3600 | 3600 | 3600 | 3600 | 3600 |
| min n. p.    | 1    | 1    | 1    | 1    | 1    | 1    | 1    | 1    | 1    | 1    |
| max n. p.    | 8    | 8    | 4    | 4    | 4    | 4    | 8    | 8    | 8    | 8    |

Table (2) Properties of users in the simulation experiments. (The abbreviation in Row 1 is based on the corresponding users properties mentioned in Figure 5)

|         | n. u. | min u. b. | max u. b. | min u. b. f. | max u. b. f. | min u. qos | max u. qos |
|---------|-------|-----------|-----------|--------------|--------------|------------|------------|
| users 1 | 10    | 20        | 100       | 16           | 512          | 2          | 10         |
| users 2 | 20    | 20        | 400       | 16           | 512          | 2          | 15         |

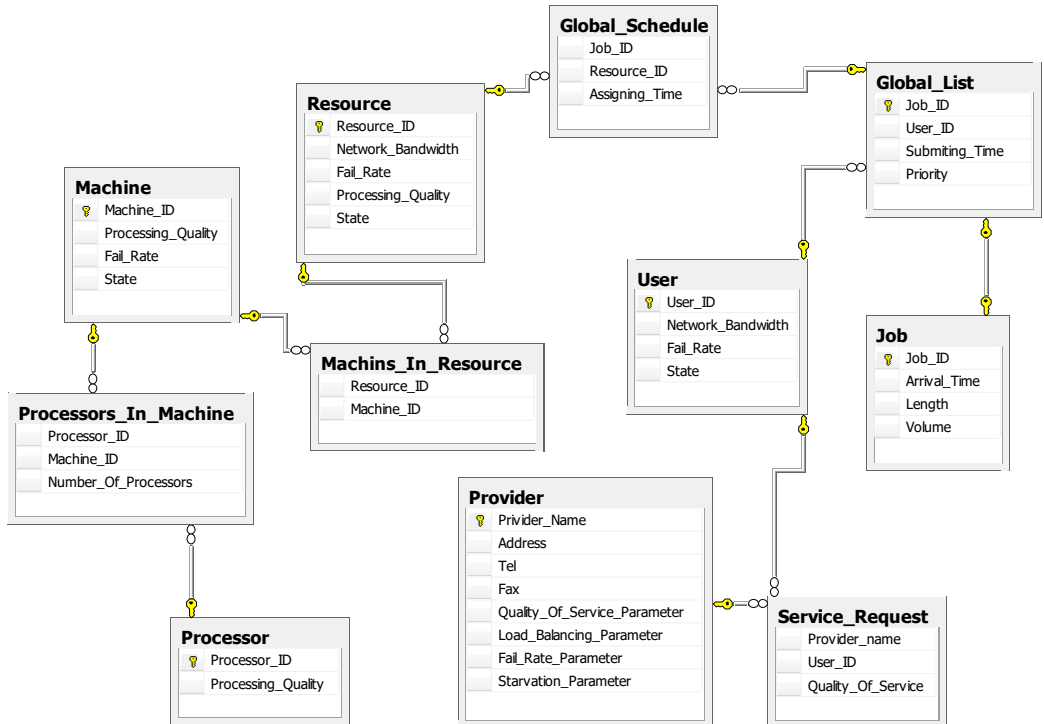

**Figure 1.** Conceptual model of the described Grid

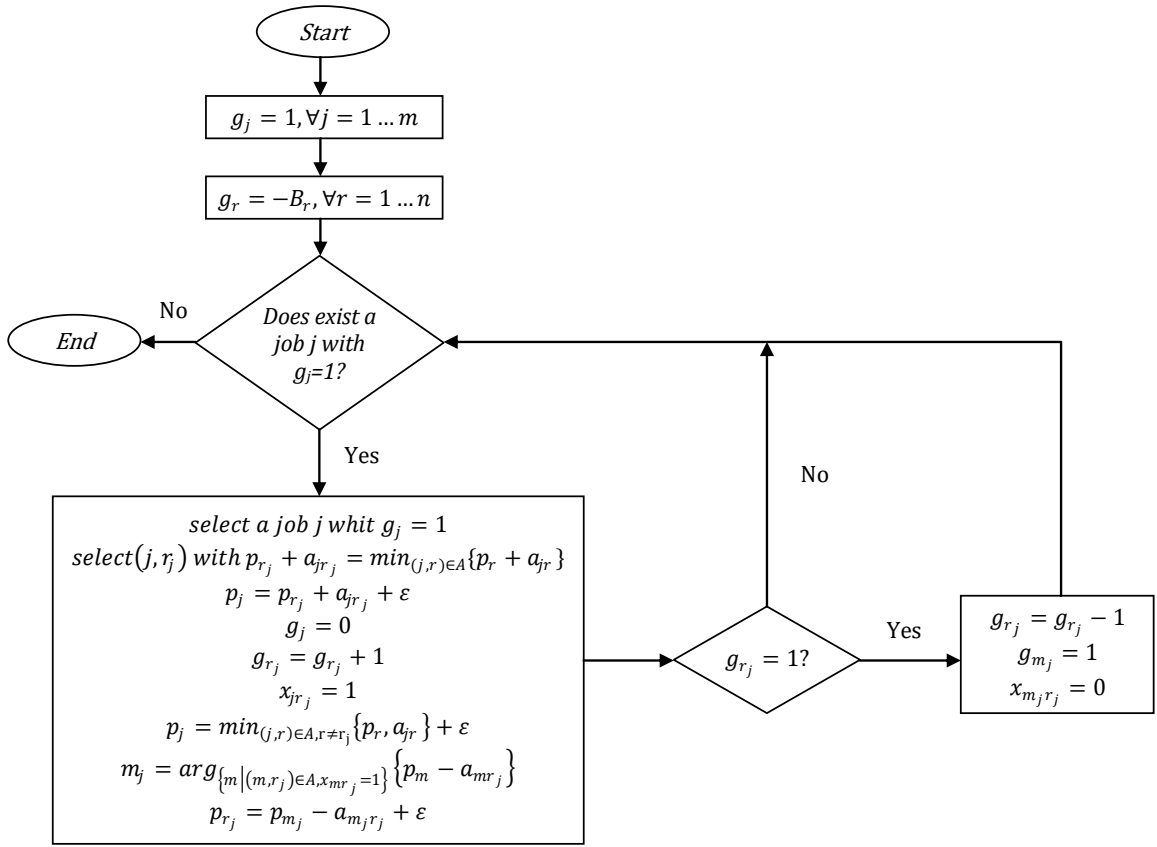

**Figure 2.** Flowchart of Auction-based heuristic for global scheduling

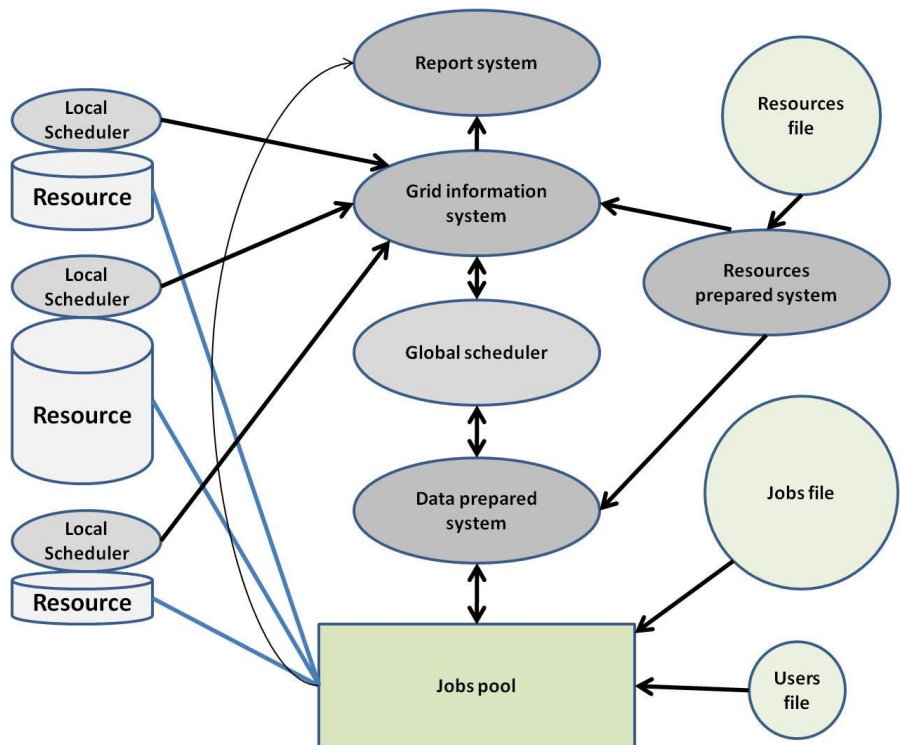

**Figure 3.** Components of Multi-S-Grid

```
grid_properties.txt

    number_of_resources=3,
    minimum_resource_bandwide=32,
    maximum_resource_bandwide=512,
    minimum_resource_bandwidth_fail_rate=30,
    max_resource_bandwidth_fail_rate=120,
    minimum_number_of_machines_in_each_resource=1,
    maximum_number_of_machines_in_each_resource=4,
    minimum_machine_fail_rate=15,
    maximum_machine_fail_rate=90,
    minimum_processor_speed=1200,
    maximum_processor_speed=3600,
    minimum_number_of_processors_in_each_machine=1,
    maximum_number_of_processors_in_each_machine=8
```

**Figure 4.**     The sample file of Grid properties

```
user_properties.txt

    number_of_users=10,
    minimum_user_bandwidth_fail_rate=20,
    maximum_user_bandwidth_fail_rate=100,
    minimum_user_bandwidth=16,
    maximum_user_bandwidth=512,
    minimum_user_quality_of_service=2,
    max_user_quality_of_service=10
```

**Figure 5.**     The sample file of user properties

```
job_properties

    minimum_job_lengh=1200,
    maximum_job_lengh=12000,
    minimum_job_input_volume=32,
    maximum_job_input_volume=1024
```

**Figure 6.**     The sample file of job properties

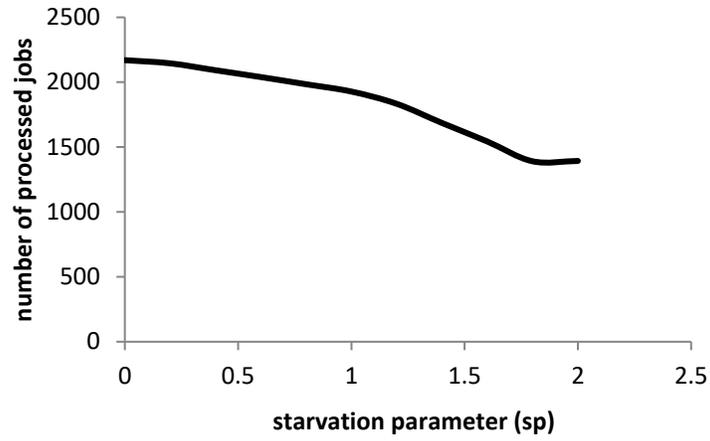

**Figure 7.** The effect of starvation parameter on the number of processed jobs.

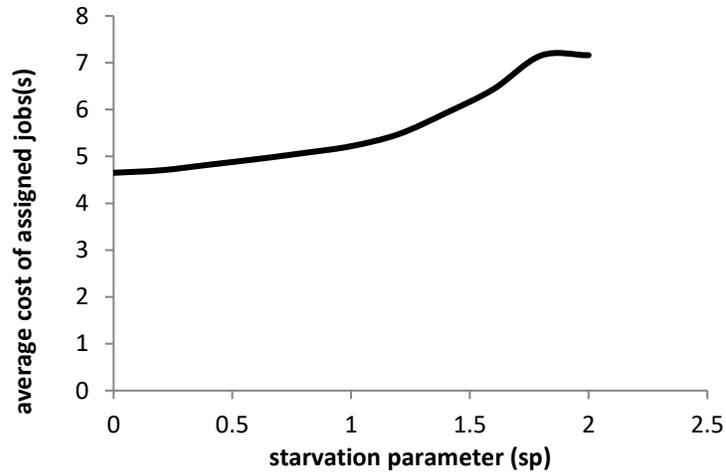

**Figure 8.** The effect of starvation parameter on the average cost of assigned jobs.

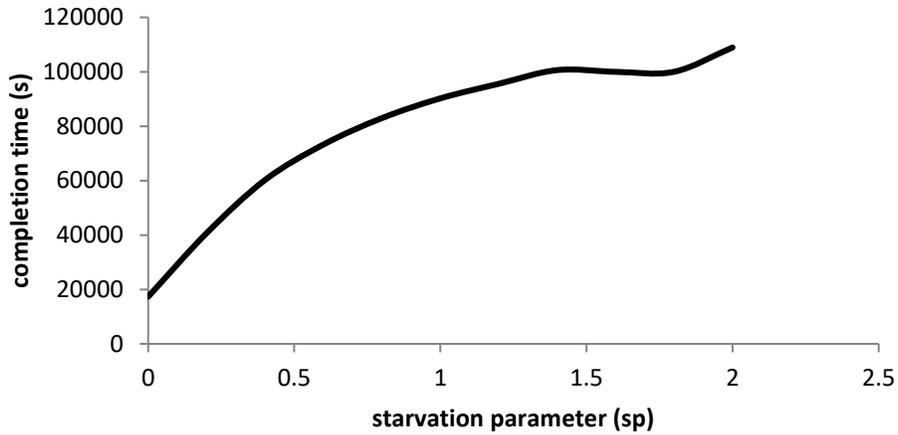

**Figure 9.** Completion time variation for different values of starvation parameter.

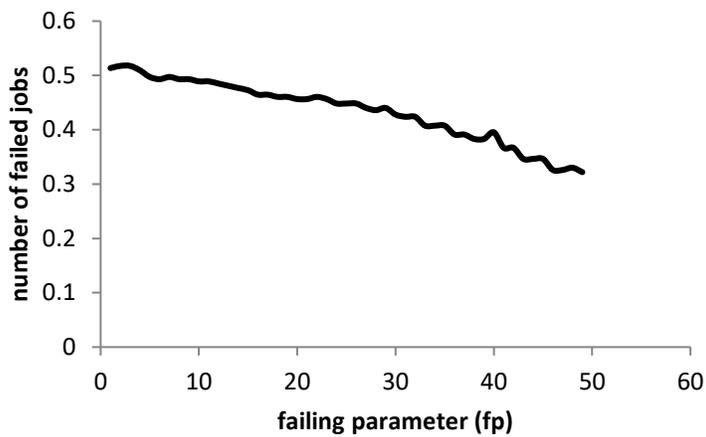

**Figure 10.** The effect of failing parameter on the number of failed jobs.

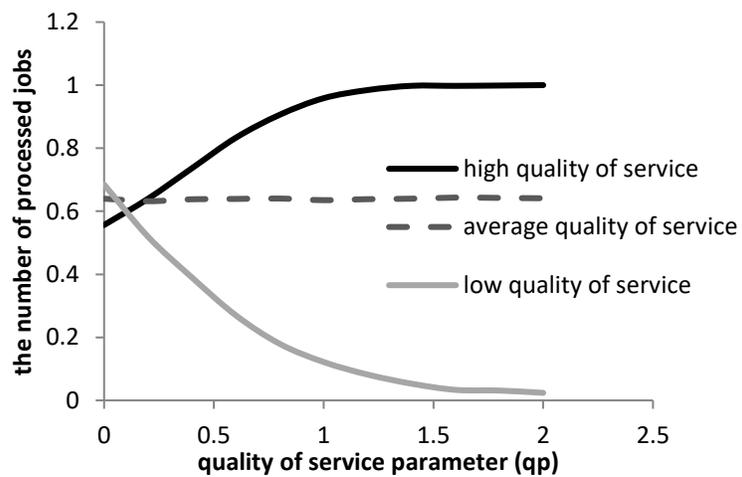

**Figure 11.** The effect of quality of service parameter on the number of processed jobs.

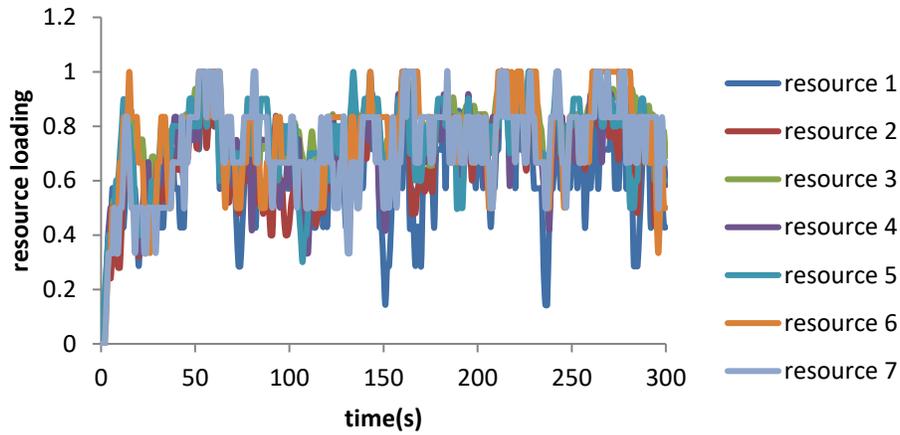

**Figure 12.** The effect of load balancing of resources in a Grid under peak

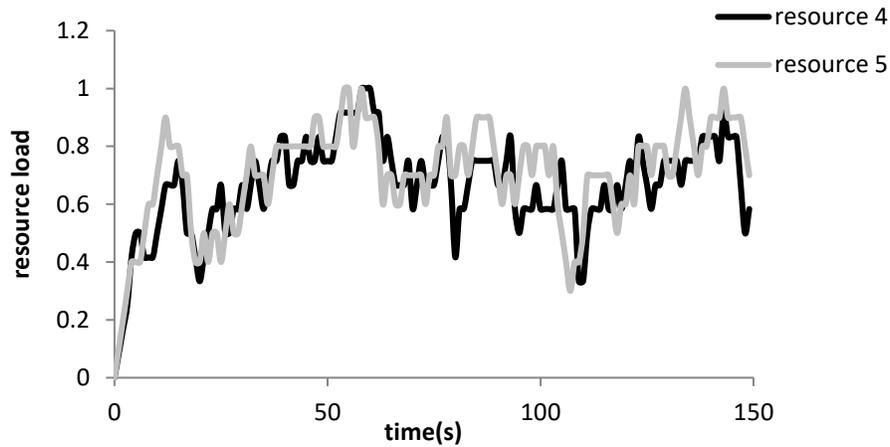

**Figure 13.** The effect of load balancing of resources in a Grid under peak (The results of two resources)

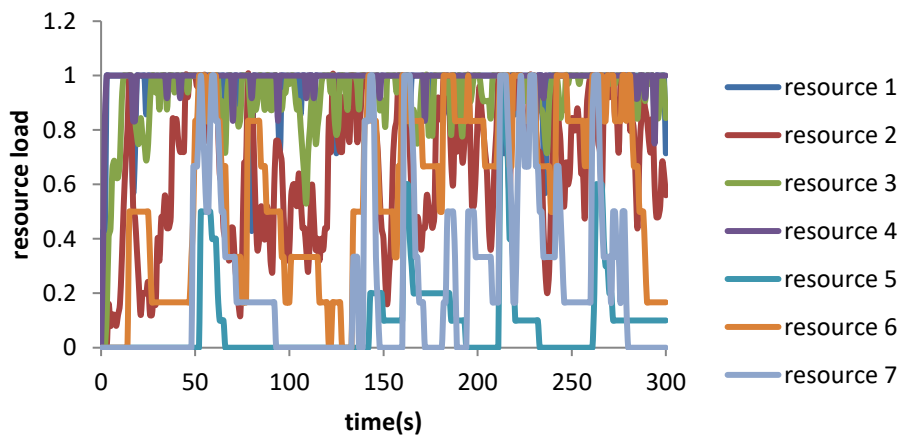

**Figure 14.** The fraction of resource loading in a Grid under peak, without load balancing strategy.

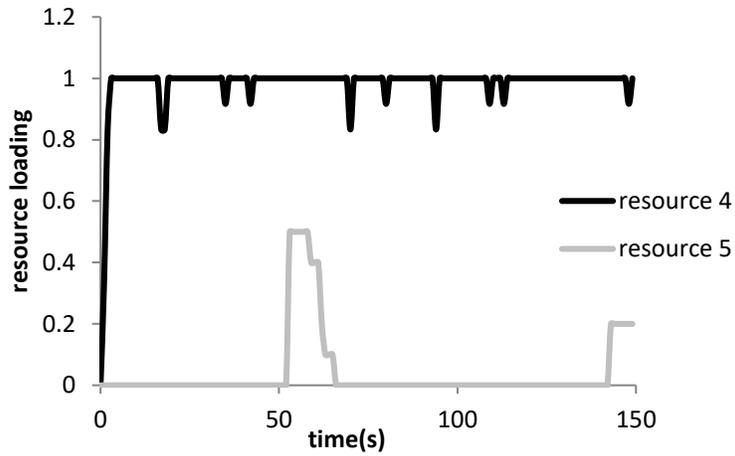

**Figure 15.** The fraction of resource loading in a Grid under peak, without load balancing strategy (The results of two resources).

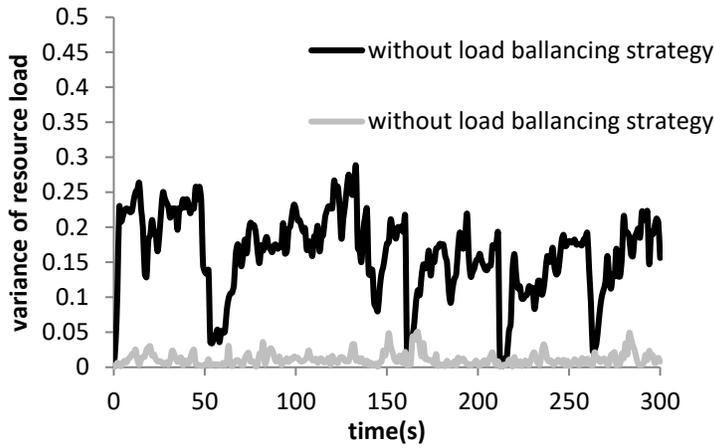

**Figure 16.** The variance of resource loading in a Grid under peak.

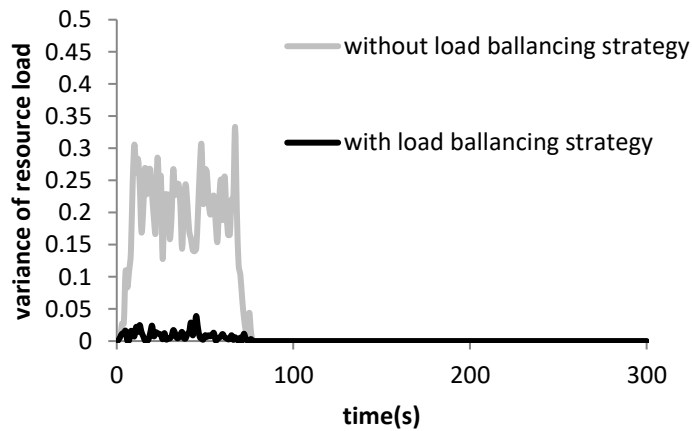

**Figure 17.** The variance of resource loading in a Grid in peak period.